\documentclass[journal]{IEEEtran}
\usepackage[english]{babel}
\usepackage[utf8]{inputenc}
\usepackage[pdftex]{graphicx}
\usepackage{amsmath,mathrsfs}
\usepackage{algorithmic}
\usepackage{array}
\usepackage{mdwmath}
\usepackage{mdwtab}
\usepackage{eqparbox}
\usepackage[tight,footnotesize]{subfigure}
\usepackage[pdftex,usenames,dvipsnames]{color}
\usepackage{siunitx}
\usepackage{listings}
\usepackage[hidelinks]{hyperref}
\usepackage{parskip}
\usepackage{color}
\usepackage{comment}
\usepackage{braket}

\definecolor{mygreen}{rgb}{0,0.6,0}
\definecolor{mygray}{rgb}{0.5,0.5,0.5}
\definecolor{mymauve}{rgb}{0.58,0,0.82}
\begin{document}

\title{Explaining Retrocausality Phenomena in Quantum Mechanics using a Modified Variational Principle}

\author{Luis Fernando Mora Mora, Eng.\\ \textit{luis.mora\_m@ucr.ac.cr} \\ \textit{lmoramora94@gmail.com} \\~\IEEEmembership{University of Costa Rica}}% 

% The paper headers
\markboth{}{}
\maketitle

\begin{abstract}
A modified lagrangian with causal and retrocausal momenta was used to derive a first \textit{causal} wave equation and a second \textit{retrocausal} wave equation using the principle of least action. The retrocausal wave function obtained through this method was found to be equivalent to the complex conjugate of the causal wave function, thus leading to the conclusion that a retrocausal effect was already implicit in quantum mechanics through the means of complex conjugation of the wave function when computing the probability density for a particle. Lastly, the same variational principle was employed with a fractionary langriangian, (that is, containing fractional Riemann derivatives) to obtain a pair of modified wave equations, one causal and other retrocausal, both of which correspond to the differential equation of a damped oscillator in the free particle (potential energy V=0) case. The solutions of this damped wave equations remain to be explored.
\end{abstract}

%\begin{IEEEkeywords}
% Esto de último
%\end{IEEEkeywords}

\IEEEpeerreviewmaketitle

% Requerimientos: relevancia y antecedentes

% Referencias - Chek

\section{Introduction}

Lagrangian mechanics started back in 1788 with Joseph-Louis Lagrange, who derived a mathematical formalism, through the means of calculus of variations, that reproduced Newton's laws from the analysis of the energies of a system \cite{taylor2005classical, goldstein2002classical,greiner2009classical}. By forcing the action of the system to be stationary, a set of equations, called the Euler-Lagrange equations, can be obtained which describe how a system evolves in time. This method provided the equations of motion, identical to the ones obtained through Newton's second law, without the consideration of force vectors but through the means of principle of ``least action''. Similarly, more that a hundred years later, using the same principle of least action and incorporating the principles of Hamilton-Jacobi theory, Erwin Schr\"{o}dinger derived his famous equation \cite{schroedinger1993quantization} for explaining energy quantization in atoms as the solution of an eigenvalue problem. From classical to quantum mechanics, Lagrange's work has taken a fundamental role in discovering the laws that govern our universe.\\

Based on the formulation of lagrangians used by Dreisigmeyer \& Young \cite{dreisigmeyer2003nonconservative} in frictional systems, this article proposes a modified lagrangian and variational principle later to be used in a similar manner as  Schr\"{o}dinger did, to obtain two different wave equations, one causal and the other one retrocausal, or backwards causal. In this work, ``anti-causal'', ``retrocausal'' and ``backwards-causal'' are all terms to be understood as the same phenomena, where time is flowing in the opposite direction or where information seems to travel backwards in time. This type of retrocausal effects have been observed in the so-called ``delayed choice'' interferometry experiments, also known as ``quantum erasers'' (see for example \cite{kim2000delayed,walborn2002double}), which are still a topic of debate. On the other hand, the term ``causal'' refers to our usual perception of time flowing in one direction from the past to the present, and where only events from the past seem to affect our present.

\subsection{Fractional Derivatives and Fractional Lagrangians}

Consider an interval of time $a<t<b$, where $a$ and $b$ represent arbitrary parameters. Now, consider a system with a general coordinate $q$ and its fractional derivative of order $\alpha$. If $m$ is the smallest integer greater than $\alpha$, the causal Riemann derivative of q \cite{herrmann2011fractional,SGil} is given by:  

\begin{equation}
    \label{eq:causalderivative}
     {}^{}_{a}D^{\alpha}_{t}[q(t)] = \frac{d^m}{dt^m}\Big[ \frac{1}{\Gamma(m-\alpha)}\int_a^t q(\tau)(t-\tau)^{m-\alpha-1}d\tau\Big]
\end{equation}

But to simplify notation, the causal derivative is written as:

  \begin{equation}\label{ar13}  
    {}^{}_{a}D^{\alpha}_{t}[q(t)] = _{a}q^{\alpha}_{t}
  \end{equation}

Similarly, the corresponding retrocausal derivative of q is given by:

\begin{equation}
    \label{eq:retrocausalderivative}
    \scriptsize{
     {}^{}_{t}D^{\alpha}_{b}[q(t)] = (-1)^m\frac{d^m}{dt^m}\Big[ \frac{1}{\Gamma(m-\alpha)}\int_t^b q(\tau)(\tau-t)^{m-\alpha-1}d\tau\Big]}
\end{equation}

But again, to simplify the notation, the retrocausal derivative is written as:

\begin{equation}\label{ar13}  
    {}^{}_{t}D^{\alpha}_{b}[q(t)] = {}^{}_{t}q^{\alpha}_{b}
  \end{equation}

These definitions of fractional derivatives are generally viewed as a convolution of q(t) with a ``potential'' $\Phi(t)_+ = t^{m-\alpha-1}$ in the causal case and $\Phi(t)_- = -t^{m-\alpha-1}$ in the retrocausal case. Equation \ref{eq:causalderivative} is a \textit{forward} convolution, because time is being integrated from a past point $a<t$ to a present moment \textit{t} and that gives it the ``causal'' part to its name. On the other hand, equation \ref{eq:retrocausalderivative} is a \textit{reverse} convolution, because time is being integrated from a future point b to a present time $t$. For this reason is to be called \textit{retrocausal}, but to account for this reversion the potential changes and a factor of $(-1)^m$ is introduced.  

Following the above notation, \cite{dreisigmeyer2003nonconservative} defines the following lagrangian, which is a function of both the causal and retrocausal fractional derivatives of position:

  \begin{equation}\label{ar14}  
    \mathcal{L}(q,_aq^{1}_t,_tq^{1}_b,_aq^{\alpha}_t,_tq^{\alpha}_b)
  \end{equation}
  
Using this lagrangian, the action of the system is made stationary by the following equation:

\begin{equation} 
\delta S = \delta  \int_{a}^{b}  \mathcal{L}(q, _aq^{1}_t, _tq^{1}_b, _aq^{\alpha}_t, _tq^{\alpha}_b) dt = 0
\end{equation}

The variation is calculated by taking an arbitrary small parameter $\eta$ so that $q\longrightarrow q + \eta$. In this way, the variation as a first order expansion is given by:

 \begin{equation}  
    \tiny{\delta  \int_{a}^{b}  \Big[\frac{\partial \mathcal{L}}{\partial q} \eta + \frac{\partial \mathcal{L}}{\partial _{a}q^{1}_{t}}\cdot _a\eta^{1}_{t} + \frac{\partial \mathcal{L}}{\partial _{a}q^{\alpha}_{t}}\cdot _a\eta^{\alpha}_{t} +
  \frac{\partial \mathcal{L}}{\partial _{t}q^{1}_{b}}\cdot _{t}\eta^{1}_{b} + \frac{\partial \mathcal{L}}{\partial _{t}q^{\alpha}_{b}}\cdot _t\eta^{\alpha}_b \Big] dt = 0}
 \end{equation}

Using the condition $\eta(a)= \eta(b)$, and integrating by parts, the fractional derivatives of $\eta$ can be replaced in the following way \cite{dreisigmeyer2003nonconservative}:

\begin{equation}
    \label{eq:byparts}
 \int_{a}^{b} \frac{\partial \mathcal{L}}{\partial {}^{}_{a}q^{\alpha}_{t}} {}^{}_{a}\eta ^{\alpha}_{t} dt  =  \int_{a}^{b} {}^{}_{t}D ^{\alpha}_{b}\Big[\frac{\partial \mathcal{L}}{\partial {}^{}_{b}q^{\alpha}_{t}} \Big ] \eta dt
\end{equation}

As equation \ref{eq:byparts} shows, the retrocausal derivative appears instead of the negative sign one is used to in integration by parts. By this, one can factorize the variation parameter $\eta$ in equation 7:

 \begin{equation} 
    \small{
   \int_{a}^{b}  \Big[\frac{\partial \mathcal{L}}{\partial q}  + _{t}D^{1}_{b} \frac{\partial \mathcal{L}}{\partial _{a}q^{1}_{t}}  + _{t}D^{\alpha}_{b} \frac{\partial \mathcal{L}}{\partial _{a}q^{\alpha}_{t}} + _{a}D^{1}_{t} \frac{\partial \mathcal{L}}{\partial _{t}q^{1}_{b}} + _{a}D^{\alpha}_{t} \frac{\partial \mathcal{L}}{\partial _{t}q^{\alpha}_{b}}  ] \eta dt = 0}
 \end{equation}
 
Because the action must be stationary for any value the parameter $\eta$ gets, the term in the parenthesis must be equal to zero. This gives the modified Euler-Lagrange equations for the system:

\begin{equation}\label{ar22}
\frac{\partial \mathcal{L}}{\partial q}  + _{t}D^{1}_{b} \frac{\partial \mathcal{L}}{\partial _{a}q^{1}_{t}}  + _{t}D^{\alpha}_{b} \frac{\partial \mathcal{L}}{\partial _{a}q^{\alpha}_{t}} + _{a}D^{1}_{t} \frac{\partial \mathcal{L}}{\partial _{t}q^{1}_{b}} + \\ _{a}D^{\alpha}_{t} \frac{\partial \mathcal{L}}{\partial _{t}q^{\alpha}_{b}}  = 0
 \end{equation} 

For example, if this Euler-Lagrange equation is applied to the following lagrangian \cite{dreisigmeyer2003nonconservative}:

\begin{equation}
    \mathcal{L} = \frac{-m}{2}( _{a}q^{1}_{t})(_{t}q^{1}_{b})-\frac{C}{2}( _{a}q^{\alpha}_{t})(_{t}q^{\alpha}_{b}) -V(q)
\end{equation}

By splitting the potential energy in two, one obtains the following equations of motion when the causal part is forced to vanish separately from the retrocausal part:

\begin{equation}\label{ar31}
m{}^{}_{a}D^{2}_{t}[q]  + C({}^{}_{a}D^{2\alpha}_{t}[q]) = -\frac{\partial V}{\partial q}
\end{equation} 

\begin{equation}\label{ar32}
m{}^{}_{t}D^{2}_{b}[q]  + C(_tD^{2 \alpha}_{b}[q]) = -\frac{\partial V}{\partial q}
\end{equation} 

If we let then $\alpha \rightarrow \frac{1}{2}$, and take the potential to be $V(q) = \frac{1}{2}m\omega^2q^2$ then the causal and retrocausal equations for a damped oscillator are obtained \cite{SGil}:

\begin{equation}
m\ddot q + C\dot q + m\omega^2q = 0 
\end{equation}

\begin{equation}
m\ddot q - C\dot q + m\omega^2q = 0 
\end{equation}

The damped oscillator is known to have three types of solution: over-damped, critically damped and under-damped. Either way, equation 14 represents a stable process where the system returns to equilibrium (q=0) after enough time has passed because the term $C\dot q$ dissipates the energy of the system as heat. On the other hand, equation 15 represents an \textit{unstable} process because the term $-C\dot q$ is not dissipating but \textit{introducing} energy to the system. However, these two equations are supposed to describe the same physical event in time with some appropiate boundary conditions. In this aspect, one is lead to conclude that the unstable retrocausal process must be interpreted as ``watching the a stable process in reverse'', as if the stable process was a movie, to play the movie backwards.    
 
\section{Modification to the Variational Principle}

In the work  \cite{dreisigmeyer2003nonconservative}, the author decides to split the potential in two in order to obtain the right equations of motion. Nonetheless, here is proposed a way in which it is not necessary to split the potential in two and which gives a more elegant formulation of the problem. Consider the following lagrangian, similar to that in equation 5, but which is also a function of their zeroth-order causal and retrocausal derivatives: 

\begin{equation}\label{ar14}  
    \mathcal{L}(_aq^{0}_t,_tq^{0}_b,_aq^{1}_t,_tq^{1}_b,_aq^{\alpha}_t,_tq^{\alpha}_b,)
  \end{equation}
 
 Following the same procedure of taking $ q\longrightarrow q + \eta$ and expanding the lagrangian to first order, the following modified euler-lagrange equations are obtained:

\begin{equation}\label{ar22}
 _{t}D^{1}_{b} \frac{\partial \mathcal{L}}{\partial _{a}q^{1}_{t}} +
 _{t}D^{\alpha}_{b} \frac{\partial \mathcal{L}}{\partial _{a}q^{\alpha}_{t}} +
_{t}D^{0}_{b} \frac{\partial \mathcal{L}}{\partial _{a}q^{0}_{t}} = 0
 \end{equation}  
 
\begin{equation}
_{a}D^{1}_{t} \frac{\partial \mathcal{L}}{\partial _{t}q^{1}_{b}} + _{a}D^{\alpha}_{t} \frac{\partial \mathcal{L}}{\partial _{t}q^{\alpha}_{b}} + _{a}D^{0}_{t} \frac{\partial \mathcal{L}}{\partial _{t}q^{0}_{b}}  = 0
\end{equation}

And analogously to equation 11, the following lagrangian is proposed:

\begin{multline}\label{ar28}
\mathcal{L} = m( _{a}q^{1}_{t})(_{t}q^{1}_{b})+C( _{a}q^{\alpha}_{t})(_{t}q^{\alpha}_{b})+k( _{a}q^{0}_{t})(_{t}q^{0}_{b})
\end{multline}

This lagrangian is different to that of equation 11 because the negative signs used before are no longer needed. Here the constant $k$ corresponds to the spring constant of the oscillator, which is exchangable with $m\omega^2$. Applying equations 17 and 18 on this lagrangian, one obtains the equivalent to equations 12 and 13, and the latter equations 13 and 14 for the damped oscillator: 

\begin{equation}\label{ar32}
m{}^{}_{t}D^{2}_{b}[q]  + C(_tD^{2 \alpha}_{b}[q]) + k(_{t}D^{0}_{b}[q]) = 0
\end{equation} 

\begin{equation}\label{ar31}
m{}^{}_{a}D^{2}_{t}[q]  + C({}^{}_{a}D^{2 \alpha}_{t}[q]) + k(_{a}D^{0}_{t}[q]) = 0
\end{equation} 

Following this reasoning, the concept of a lagrangian can be reduced to a sum over the products of causal and retrocausal, $\beta$-order derivatives in the following form:

\begin{equation}
   \mathcal{L} = \sum_\beta C_{\beta}( _{a}q^{\beta}_{t})(_{t}q^{\beta}_{b})
\end{equation}

Here, $C_\beta$ is some real coefficient, and $\beta$ can be any real number. For the lagrangian in equation 19, $\beta$ would be summing over the values of 1, $\frac{1}{2}$ and 0, and the coefficients $C_\beta$ are \textit{m}, \textit{C} and \textit{k} respectively. This generalization of lagrangian can then yield any pair of causal and retrocausal equations of motion of any given order, according to the $\beta$ parameter used. This generalizes the Euler-Lagrange equations to the following form:

\begin{equation}
      _{a}D^{\beta}_{t} \frac{\partial \mathcal{L}}{\partial _tq^{\beta}_{b}} = 0
\end{equation}

\begin{equation}
  _tD^{\beta}_{b} \frac{\partial \mathcal{L}}{\partial _{a}q^{\beta}_{t}} = 0 
\end{equation}

Where there is a sum over the index $\beta$ in both equations. With these results in mind, next we proceed to analize the quantum mechanical formalism. 

\section{Quantization of Energy as an Eigenvalue Problem Revisited}

The problem of energy quantization as an eigenvalue problem was formulated by Schr\"{o}dinger in the following manner \cite{schroedinger1993quantization}:

\begin{equation}
    H(q,\frac{\partial S}{\partial q}) = E
\end{equation}

Here, according to Hamilton-Jacobi theory, the momenta in the Hamiltonian are replaced by the partial derivatives with respect to coordinates of the action \textit{S} \cite{greiner2009classical}. Now we set the goal of deriving the same formulation using the causal/retrocausal approach. Using the Legendre transform, it is possible to prove \cite{HamiltronFractional} that for the following lagrangian: 

\begin{equation}
    \mathcal{L} = m(_aq^1_t)(_tq^1_b)-V(q)
\end{equation}

The corresponding hamiltonian function is of the form:

\begin{equation}
    \mathcal{H} = \frac{1}{m}p_+p_- + V(q) = E
\end{equation}

Where the momenta $p_+$ and $p_-$ correspond to the first order causal and retrocausal derivatives in the lagrangian of equation 25. These can be also thought as \textit{causal speed} and \textit{retrocausal speed} with their corresponding \textit{causal momentum} $p_+$ and \textit{retrocausal momentum} $p_-$. Under this formalism and following the steps of Schr\"{o}dinger  \cite{schroedinger1993quantization}, the following causal and retrocausal actions are proposed respectively as:

\begin{equation}
    \mathcal{S}_+ = K\ln(\psi_+) =  \frac{\hbar}{\sqrt{2}}\ln(\psi_+)
\end{equation}
\begin{equation}
    \mathcal{S}_- = K\ln(\psi_-) =  \frac{\hbar}{\sqrt{2}}\ln(\psi_-)
\end{equation}

Where K is a constant with action units. In the above equations, $\psi_+$ corresponds to a \textit{causal wave function}, and similarly $\psi_-$ corresponds to a \textit{retrocausal wave function}. By doing this, the momenta $p_+$ and $p_-$ can be replaced by the Hamilton-Jacobi transformation:

\begin{equation}
    p_+ = \frac{\partial \mathcal{S}_+}{\partial q} = \frac{\hbar}{\sqrt{2}\psi_+}\frac{\partial \psi_+}{\partial q}
\end{equation}

\begin{equation}
    p_- = \frac{\partial \mathcal{S}_-}{\partial q} = \frac{\hbar}{\sqrt{2}\psi_-}\frac{\partial \psi_-}{\partial q}
\end{equation}

So when the momenta are substituted in the Hamiltonian of equation 26, one obtains:

\begin{equation}
    \frac{\hbar^2}{2m}\nabla \psi_+ \nabla \psi_- - (E-V(q))\psi_+ \psi_- = 0
\end{equation}

Analogously to Schr\"{o}dinger \cite{schroedinger1993quantization}, the variational problem for minimizing the energy is formulated as:

\begin{equation}
\label{ec:mixedenergy}
    \delta \int \Big[ \frac{\hbar^2}{2m}\nabla \psi_+ \nabla \psi_- + (V(q)-E)\psi_+ \psi_-\Big]dxdydz = 0
\end{equation}
Where the integration is being taken over all space. For this integral to be stationary it is necessary to have:

\begin{equation}
    \nabla^2\psi_+ + \frac{2m}{\hbar^2}(E-V)\psi_+ = 0
\end{equation}

\begin{equation}
    \nabla^2\psi_- + \frac{2m}{\hbar^2}(E-V)\psi_- = 0
\end{equation}

Thus, a wave equation is obtained for $\psi_+$ separately from a wave equation for $\psi_-$. Since both equations where obtained from the same parameter of energy E, this means that both $\psi_+$ and $\psi_-$ must have the \textit{same} energy eigenvalue. When solved in the energy basis and going to the usual bra-ket notation, both wavefunctions can be expressed as \cite{shankar2012principles}:

\begin{equation}
    \ket{\psi_+} = \sum_E \ket{E}\bra{E}e^{\frac{-iEt}{\hbar}}
\end{equation}

Now, the retrocausal wave function $\psi_-$ has the same eigenfunction solution as $\psi_+$ but experiences a temporal inversion, analogously to the temporal inversion the backwards-causal damped oscillator of equation 15 experiences. For this reason, the exponent with the temporal dependence should have a minus sign, implying time reversion:

\begin{equation}
    \ket{\psi_-} = \sum_E \ket{E}\bra{E}e^{\frac{iEt}{\hbar}}
\end{equation}

 Thus, by measuring the state of one of the wave functions, the other is forced to colapse to the same state, because they share the same energy eigenvalue. In other words, the \textit{causal} wave function collapses the \textit{retrocausal} wave function, and viceversa. Equation 37 looks very familiar in quantum mechanical terms, since $\psi_-$ is the complex conjugate of the causal wave function $\psi_+$. This shows there is an equivalence in taking the complex conjugate of a wave function as taking it as a retrocausal wave function produced by a retrocausal action. Then, the probability density for finding the particle at position x can be  obtained from:

\begin{equation}
    \rho(x) = \psi_+(x)\psi_-(x)
\end{equation}

This being equivalent to the familiar probability amplitude:

\begin{equation}
    \rho(x) = |\psi(x)|^2 = \psi(x)\psi(x)^*
\end{equation}

Thus, the retrocausality action taken into consideration is already implicit in quantum theory, hiding in the complex conjugation of the classical wave function, and this should be no surprise because complex conjugation implies time reversal in time symmetry analysis. \cite{shankar2012principles}. 

\section{Extension to the Fractional Case}

Now lets consider the possibility of applying a fractional lagrangian to quantum theory by considering a variation of the lagrangian in equation 19, which models dissipative systems, by taking a general potencial V:

\begin{equation}
    \mathcal{L} = m(_aq^1_t)(_tq^1_b)+B(_aq^{\frac{1}{2}}_t)(_tq^{\frac{1}{2}}_b)-V(q)
\end{equation}

Here B is again a damping coefficient, not to be confused with $c$, the speed of light. For this lagrangian, the corresponding Hamiltonian function is\cite{HamiltronFractional}:

\begin{equation}
    \mathcal{H} = \frac{1}{m}p_+p_- + \frac{1}{B}p^{\frac{1}{2}}_+p^{\frac{1}{2}}_- + V(q) = E
\end{equation}

And as you might have guessed, the index $\frac{1}{2}$ over the momenta refers to the corresponding half derivatives in the lagrangian of equation 40, not to be confused with square root. This means a \textit{half-causal momentum} $p^{\frac{1}{2}}_+$ and a \textit{half-retrocausal momentum} $p^{\frac{1}{2}}_-$ are introduced to the Hamiltonian. In this case, the following substitution is proposed:

\begin{equation}
    p^{\frac{1}{2}}_+ = \frac{\partial^{\frac{1}{2}} \mathcal{S}_+}{\partial q^{\frac{1}{2}}} = \frac{(\hbar mc)^{1/2}}{\sqrt{2}\psi_+}\frac{\partial^{\frac{1}{2}} \psi_+}{\partial q^{\frac{1}{2}}}
\end{equation}

\begin{equation}
    p^{\frac{1}{2}}_- = \frac{\partial^{\frac{1}{2}} \mathcal{S}_-}{\partial q^{\frac{1}{2}}} = \frac{(\hbar mc)^{1/2}}{\sqrt{2}\psi_-}\frac{\partial^{\frac{1}{2}} \psi_-}{\partial q^{\frac{1}{2}}}
\end{equation}

Where the factor $(\hbar mc)^{1/2}$ ensures the correct momentum units \cite{herrmann2011fractional}. Thus one obtains the following quantity for the energy variation:

\begin{multline}
   \delta \int_a^b \Big[\frac{\hbar^2}{2m}\nabla \psi_+ \nabla \psi_- + \frac{\hbar mc}{2B}\nabla^{\frac{1}{2}} \psi_+\nabla^{\frac{1}{2}} \psi_- \\- (E-V(q))\psi_+ \psi_-\Big]dxdydz = 0 
\end{multline}
   
For this integral to be stationary, it is necessary that:

\begin{equation}
    \frac{\hbar^2}{2m}\nabla^2\psi_+ + \frac{\hbar mc}{2B}\nabla^{(1/2)^2}\psi_+ + (E-V(q))\psi_+ = 0
\end{equation}

\begin{equation}
    \frac{\hbar^2}{2m}\nabla^2\psi_- + \frac{\hbar mc}{2B}\nabla^{(1/2)^2}\psi_- + (E-V(q))\psi_+ = 0
\end{equation}

These two can further be reduced by multiplying on both sides by $\frac{2m}{\hbar^2}$ to obtain:

\begin{equation}
    \nabla^2\psi_+ + \frac{m^2c}{\hbar B}\nabla^{(1/2)^2}\psi_+ + \frac{2m}{\hbar^2}(E-V(q))\psi_+ = 0
\end{equation}

\begin{equation}
\nabla^2\psi_- + \frac{m^2c}{\hbar B}\nabla^{(1/2)^2}\psi_- + \frac{2m}{\hbar^2}(E-V(q))\psi_+ = 0
\end{equation}

Or these two can be written in the usual damped oscillator way with a $\xi$ damping factor:

\begin{equation}
    \nabla^2\psi_+ + 2\xi\nabla^{(1/2)^2}\psi_+ + \frac{2m}{\hbar^2}(E-V(q))\psi_+ = 0
\end{equation}

\begin{equation}
\nabla^2\psi_- + 2\xi\nabla^{(1/2)^2}\psi_- + \frac{2m}{\hbar^2}(E-V(q))\psi_+ = 0
\end{equation}

Where $\xi = m^2c/2\hbar B$. Note that the fractional del operator of order 1/2 when squared satisfies that:

\begin{equation}
    \nabla^{(1/2)^2} = (\frac{\partial^{(1/2)}}{\partial x^{(1/2)}},\frac{\partial^{(1/2)}}{\partial y^{(1/2)}},\frac{\partial^{(1/2)}}{\partial z^{(1/2)}})^2
\end{equation}

\begin{equation}
    \nabla^{(1/2)^2} = \frac{\partial}{\partial x} + \frac{\partial}{\partial y} + \frac{\partial}{\partial z} = \nabla\cdot(\hat{x},\hat{y},\hat{z})
\end{equation}

If the 1-dimensional case where V=0 is taken (free particle), equations 49 and 50 reduce to a \textit{damped oscillator differential equation} analogous to equation 14:

\begin{equation}
    \frac{d^2\psi_+}{dx^2}+ 2\xi\frac{d\psi_+}{dx} + \frac{2mE}{\hbar^2}\psi_+ = 0
\end{equation}

\begin{equation}
    \frac{d^2\psi_-}{dx^2}+ 2\xi\frac{d\psi_-}{dx} + \frac{2mE}{\hbar^2}\psi_- = 0
\end{equation}

Note that the quantum retrocausal case is not an unstable system, in contrast with the classical retrocausal response that equation 15 shows. So, by using the principle of least action with fractional derivatives, a fractionary or ``damped'' wave equation can be obtained from a lagrangian which models dissipative systems using fractional derivatives. The solutions of this type of damped wave equation could present a simpler alternative to the already existing fractionary wave equation developed by Laskin \cite{laskin2000fractional}:

\begin{equation}
    \nabla^\alpha\psi + \big(\frac{2m}{\hbar^2}\big)^\alpha(E-V(q))\psi = 0 
\end{equation}

whose solutions are given in terms of the Fox H-Function. It should be said that fractionary equations of this type, which include fractional operators, have been linked to several interesting and strange phenomena, such as anomalous diffusion in plasmas \cite{corkum1973anomalous,drummond1962anomalous} and quark confinement \cite{herrmann2011fractional}.

\section{Conclusions}

When considering a non-dissipative system, a retrocausal action and a retrocausal wave function was found to be equivalent to the complex conjugate of the wave function described in classical quantum theory. Having started from a causal and retrocausal formulation, a probability density was obtained and was found to be equivalent to the ``original'' quantum theory probability density. This may imply that retrocausality is already implicit in quantum theory, and information from the future is brought to the present by the complex conjugate of the wave function. The effect of a retrocausal wave function remains to be applied and analyzed in quantum eraser experiments to name a possible application. \\

The theory here developed should be able to be expanded to systems of many particles simply by changing the complex conjugates of wave functions for retrocausal wave functions. This means that interference terms describing multi-particle systems could be interpreted as interference from the causal wave function with the retrocausal wave function (a sort of past-future interference).\\

On the other hand, when considering a system with energy dissipation, modelled by a fractional lagrangian, a modified version of the time-independent wave equation was obtained, which corresponds to a damped oscillator in the free particle case, showing a great similarity with the damped oscillations caused by friction in greater scale mechanical or electrical systems. The solutions of this ``damped'' wave equation remain to be explored in further works.

\bibliographystyle{IEEEtran}
\bibliography{bibliography}

\end{document}